\documentclass[aps,amsmath,amssymb,notitlepage,noeprint,nofootinbib,superscriptaddress]{revtex4-1}
\pdfoutput=1
\usepackage[latin9]{inputenc}
\usepackage{braket}
\usepackage{xcolor}
\usepackage{verbatim}
\usepackage{graphicx}
\usepackage{esint}
\usepackage{epstopdf}
\usepackage[colorlinks,linkcolor=purple,citecolor=green!50!black,urlcolor=blue!70!black]{hyperref}
\hypersetup{pdfauthor={Facoetti et al.},pdftitle={Classical glasses, black holes, and strange quantum liquids}}

\DeclareMathOperator{\sign}{sgn}

\usepackage{bbm}

\begin{document}

\title{The relation between Parisi scheme and  multi-thermalized  dynamics
in finite dimensions}

\author{Silvio Franz }
\affiliation{LPTMS, UMR 8626, CNRS, Univ. Paris-Sud,
  Universit\'e Paris-Saclay, 91405 Orsay, France}
  \author{J Kurchan}
\affiliation{Laboratoire de Physique de l'Ecole normale sup\'erieure, ENS, Universit\'e PSL, CNRS, 
Sorbonne Universit\'e, Universit\'e de Paris, F-75005 Paris, France}

~\\~\\

\begin{abstract}  
In this note we summarize the connections between equilibrium and slow out of equilibrium dynamics in finite dimensional glasses, such as we understand them today.
If we assume that a finite-dimensional system {is stable with respect to a family of weak random perturbations } (stochastic stability), then  its  dynamics have a  `Multithermalization' structure if and only if the Boltzmann-Gibbs distribution obeys an Ultrametric Parisi distribution. 

\end{abstract}
\maketitle
\tableofcontents

\section{  Introduction}

A 
system whose equilibrium measure follows the Parisi scheme \cite{MezardParisiVirasoro},
commonly known as Replica Symmetry Breaking (RSB), 
will take an infinite time to reach its equilibrium starting form random configurations, a process called `aging'. 
It may also be driven into an out of equilibrium steady-state by an infinitesimal drive, such as shear \cite{thalmann2001aging,berthier2000two}, or time-dependence of disorder  \cite{horner1992dynamics}. 
If the relaxation times are long, or, in  a steady-state if the drive is weak, the dynamics are slow: this is the regime we are interested in.

The out of equilibrium dynamics under these circumstances is  a very specific one \cite{sompolinsky1982relaxational,sompolinsky1981dynamic,horner1992dynamics,CugliandoloKurchan,cugliandolo1994out,franz1994off}. At given times one
may define an effective temperature with a thermodynamic meaning \cite{CugliandoloKurchanPeliti}, it is the same for all observables. 
Different temperatures are possible, but in different `time scales', a notion one has to define. We refer to this situation as `multithermalization' \cite{contucci2019equilibrium,contucci2020stationarization}.

In finite dimensional systems, neither the Parisi scheme nor the 
multi-thermalization scenario are proven to hold. However, the Parisi kind of ordering appears as the only possible non ergodic state which is 
stable  against small random perturbations in the Hamiltonian.  




Moreover, it turns out that Parisi RSB and multi-thermalization 
are deeply related: one
of them holds if and only if the other  also does.  
The temperatures involved in the slow dynamics then relate to the average overlap distribution computed for equilibrium in the Parisi scheme.
This double implication may be argued on the basis of a strategy devised by Franz, M\'ezard, Parisi and Peliti (FMPP) \cite{franz1998measuring,franz1999response}
and later generalized in \cite{kurchan2021time} 
{ that uses linear response theory and equilibration of bulk quantities to compare dynamical quantities to equilibrium one.
}
An extra element that is of central importance for the validity of both schemes, as we now realize, is the fact that the slow glassy dynamics  is {\em time-reparametrization soft}: small perturbations dramatically stretch the timescales. Given the relation described above between static replica theory and  slow dynamics, it is natural to search for reparametrization invariances in the Parisi replica context,
and indeed questions like these have appeared in the literature. The whole matter deserves further clarification.

\section{Equilibrium: Ergodicity Breaking and the Parisi scheme}

Replica Symmetry Breaking is ubiquitous in disordered systems  
with long range interactions \cite{MezardParisiVirasoro}.
Its {successful implementation}  by Giorgio Parisi dates back to the study of the 
Sherrington-Kirkpatrick model of a Spin Glass 
\cite{sherrington1975solvable}, in the effort of 
parametrizing a $n\times n$ matrix for $n\to 0$ \cite{parisi1979infinite}. 
It describes the low temperature 
disordered frozen state with broken ergodicity in systems where ordinary ordering 
is not possible. For large samples, the Gibbs measure is dominated by a few 
'valleys' unrelated by symmetry and in random positions in configuration space. 
The distances between these valleys have broad distributions, 
but they are strongly constrained, in fact only isosceles  triangles, with the different side the smallest   are possible in the thermodynamic limit, the property of `ultrametricity'. 
The mathematical analysis of long range systems 
\cite{talagrand2003spin,panchenko2013sherrington} 
has fully confirmed 
and considerably enriched the picture coming from the physicist's 
analysis based on 
the iconoclastic replica method and the more conventional, but still non rigorous 
Cavity Method. The Parisi picture is also a natural candidate to describe ideal thermodynamic glassy 
phases of systems with short range interactions. However, its 
validity in three and more in general in any finite dimensional systems has been seriously questioned \cite{fisher1986ordered,bray1987chaotic}. 
The best available numerical simulations give  strong indications in favor of RSB above a critical dimension that is somewhere above two and below four.\\
The least one can say is that RSB provides a very good statistical description of samples of finite sizes, or, equivalently, of local properties of infinite samples. Local RSB has indeed been proven to hold in models with large but finite interaction range \cite{franz2004finite}. 
But we do not know, with a convincing level of theoretical confidence, and 
even less of mathematical rigour, if RSB really survives as long range order in the thermodynamic limit. What indeed we know, is that under some rather mild 
assumption (see later), if random broken ergodicity holds, the structure of the probability measure on random Gibbs measures, also called metastate, should be qualitatively described by the RSB scenario. In fact, the mathematicians were able to prove  that under the hypothesis that the overlap distributions verify the so called Ghirlanda-Guerra identities (GGI) \cite{ghirlanda1998general} then RSB ultrametric statistics follows \cite{panchenko2013parisi}. 
In turn, the GGE follow from the stability of Gibbs measure against small perturbation \cite{ghirlanda1998general,aizenman1998stability}, an hypothesis that holds true in mean-field and, we will argue, can be generically accepted in finite dimension.
Within Mean-Field, it soon appeared that the dynamics of RSB systems presents anomalies with respect to the one in ordinary ergodic systems \cite{sompolinsky1981dynamic,sompolinsky1982relaxational}. Recognizing the full fledged off-equilibrium character of these anomalies led to the surprise that aging phenomena, which are commonly observed in laboratory glasses, also appeared in mean-field models, giving rise to a coherent and detailed theory of dynamic ergodicity breaking \cite{CugliandoloKurchan,cugliandolo1994out}.
This is obtained by the asymptotic analysis of exact Dynamical Mean Field Theory (DMFT) equations, which display a low temperature solution violating basic equilibrium properties 
as time translation invariance and the relations between responses and correlation implied by the fluctuation dissipation theorem.  It soon appeared that RSB and aging are intertwined phenomena in mean-field. Two classes of models were identified: p-spin like models and SK like models \cite{CugliandoloKurchan,cugliandolo1994out}. In the former, one observes a decoupling between equilibrium and dynamics. 
The low temperature equilibrium measure is typically dominated by metastable states (ergodic components) with high stability, while the dynamics starting from a random configuration, tends to a marginal manifold which has higher energy density respect to the equilibrium states. In the latter, instead, the states dominating equilibrium are marginally stable, and the asymptotic dynamical solution predicts asymptotic values of extensive observables coinciding with the equilibrium ones. Unfortunately, the results of asymptotic dynamics have been rather elusive to rigorous mathematical analysis. While Parisi picture for equilibrium has been made fully rigurous by mathematicians, not much has been done in dynamics, where the only progress has been the proof of the validity of the physicist's effective one body dynamical equations and their self-consistent Dynamical Mean Field Theory closure \cite{arous1995large,ben2006cugliandolo}. As far as finite dimension is concerned the situation is similar to the one of equilibrium: the Cugliandolo-Kurchan dynamical scenario provides an excellent description of finite time spin-glass dynamics \cite{barrat1999fluctuation,berthier1999response}, but we cannot exclude that the aging responses eventually become trivial and the system crosses over to complete thermalization. 

\subsection{Overlap Statistics}

In the physics of the disordered systems in equilibrium a crucial role 
is played by the statistics of the overlap between replicas induced by the 
thermal and disorder fluctuations \cite{MezardParisiVirasoro}.  What we call replicas here are configurations extracted independently from the Boltzmann measure for the same disorder.  The overlap $Q[S,S']$ between two configurations is usually defined as $Q[S,S']=\frac{1}{N}\sum_{i=1}^N S_i S_i'$.

A first characterization of the statistics of the overlap is given by its 'average histogram' 
\begin{eqnarray}
 P(q)=\mathbb{E}\left(
 \langle \delta(Q[S_1,S_2]-q))\rangle_{12}
 \right)
\end{eqnarray}  
where we denoted as $\langle\cdot\rangle_{12}$ the average over the (product) Gibbs measure of the two replicas, and 
with $\mathbb{E}$ the average over their common disorder. It was shown by Parisi a long time ago, that a non trivial 
$P(q)$ (i.e. different from a single $\delta$ function) is associated to breakdown of ergodicity \cite{parisi1983order}. This is 
trivially true for example in systems with symmetry breaking in absence of a source, where it just reflects that the measure is uniform over  the different possible phases of the system related to one another by symmetry. 
In that case, the function can be simply collapsed to a $\delta$ by the introduction 
of an infinitesimal symmetry-breaking source that selects 
one among the possible phases, and the function $P(q)$ does not 
play any physically important role in this case.  In disordered systems $P(q)$ is not related 
to a any symmetry, and the situation can be much more interesting: a small perturbation unrelated to the original Hamiltonian cannot be expected to project on a single state, and modulo possible removal of 
accidental degeneracies of the original Hamiltonian, the function $P(q)$ of a perturbed system should stay close to the same function in the unperturbed one. 
This has deep consequences, since the multiplicity of states will affect the response properties of the system and $P(q)$ can be interpreted as the generating function of the responses to a family of random perturbations. 

\subsection{The method of random perturbations}

The study of the overlap statistics requires access to microscopic configurations. This is of course 
 easy in computer studies, but is more problematic in experiments. Fortunately one can use 
linear response theory (LRT) to give a different characterization of 
$P(q)$, as the generating function of a families of 
susceptibilities. Starting from a model with Hamiltonian $H$, one considers a perturbed system where 
random $p-spin$ like Hamiltonian are added to the energy of the system
\begin{eqnarray}
\label{pert}
H_{pert}[S]=H[S]+\sum_{p=1}^\infty \epsilon_p H_p[S]
\end{eqnarray}
\footnote{The $\epsilon_p$ have to decrease fast enough with $p$ in order 
for the system to be well defined.}
where the perturbing terms are defined as 
\begin{eqnarray}
 H_p[S]=-\sum_{i_1,...,i_p}J_{i_1,...,i_p}S_{i_1}...S_{i_p}
\end{eqnarray}
The couplings $J_{1_i,...,i_p}$ are usually taken as centered i.i.d. Gaussian variables
scaled as $\mathbb{E}\left( J_{1_i,...,i_p}^2\right)=\frac{1}{N^{p-1}}$. If the parameters 
$\epsilon_p$ are chosen to be independent of $N$, the perturbing Hamiltonian involves extensive energy in equilibrium; it is straightforward to see that 
\begin{eqnarray}
\label{ip}
I^{(p)}\equiv \frac 1 N \mathbb{E}[\langle H_p\rangle] &=& T \frac{\partial}{\partial \epsilon_p}\log Z=N \beta \epsilon_p (1-\langle q^p\rangle)\\
 \langle q^p\rangle&=&\int dq\; P_\epsilon(q) q^p
\end{eqnarray}
that can be readily obtained by integration by parts. Values of $\epsilon_p$ vanishing 
with $N$, namely $\epsilon_p\sim N^{-a}$ with $0<a<1$, have also been considered, where (\ref{ip}) continues to hold, but the energy absorbed is sub-extensive. This suggests we define 
{\it Stochastic Stability} as the property that $P_\epsilon(q)\to P_0(q)$ (modulo symmetry or accidental degeneracy of the unperturbed Hamiltonian). 

The reader may object that in an experimental system the susceptibilities defined by (\ref{ip}) 
are not more accessible than the microscopic configurations. The crucial point that 
we discuss later in the paper, however, 
is that,  in turn,  these quantities can be related to {\it dynamical} susceptibilities 
whose generator function can be in principle measured through response experiments in ageing dynamics.

\subsection{Self-averageness and higher order overlap statistics}

Higher order  statistics of the overlaps, where one considers those between more than 
two replicas, are also important. Their distributions appears when formally considering the quadratic 
fluctuations of the 
perturbing terms $H_p$. 
For example, it is easy to show that 
the average square Hamiltonian $\mathbb{E}(\langle H_p[S]^2\rangle)$ is related to the statistics of 
two overlaps $P(q_{12},q_{13})$ 
\begin{eqnarray}
 \mathbb{E}(\langle H_p[S]^2\rangle)=\epsilon_p^2\beta^2 N^2
 \left(
 1-2\langle q^p \rangle -\langle q^{2p} \rangle +\langle q_{12}^p q_{13}^p \rangle 
 \right). 
\end{eqnarray}
But $H$ is a self-averaging quantity, and at order $N^2$ 
$\mathbb{E}(\langle H_p[S]^2\rangle)=(\mathbb{E}(\langle H_p[S]\rangle))^2$. This implies a strong constraint 
on $P(q_{12},q_{13})$ which needs to verify 
\begin{eqnarray}
\label{GG1}
 P(q_{13},q_{23})=\frac{1}{2}\delta(q_{13}-q_{23})P(q_{13})+\frac{1}{2}P(q_{12})P(q_{13}),
\end{eqnarray}
A similar analysis can be applied to $\mathbb{E}(\langle H_p[S]\rangle^2)$, resulting in 
\begin{eqnarray}
\label{GG2}
 P(q_{12},q_{34})=\frac{1}{3}\delta(q_{12}-q_{34})P(q_{12})+\frac{2}{3}P(q_{12})P(q_{34}). 
\end{eqnarray}
Notice that $P(q_{12},q_{34})$ represents the average over disorder of the product of the 
disorder dependent overlap distribution of four independent replicas. 
Equation (\ref{GG2}) tells us that this is different from $P(q_{12})P(q_{34})$: 
self-averageness of the energy, from which (\ref{GG2}) is derived implies that the overlap probability
(unless trivial) is instead non-self averaging. 
Also notice that once $P(q)$ is know, the higher order 
statistics of two overlaps are fixed.  

\subsection{Triangles}
\begin{figure}
    \begin{center}
\includegraphics[width=8cm]{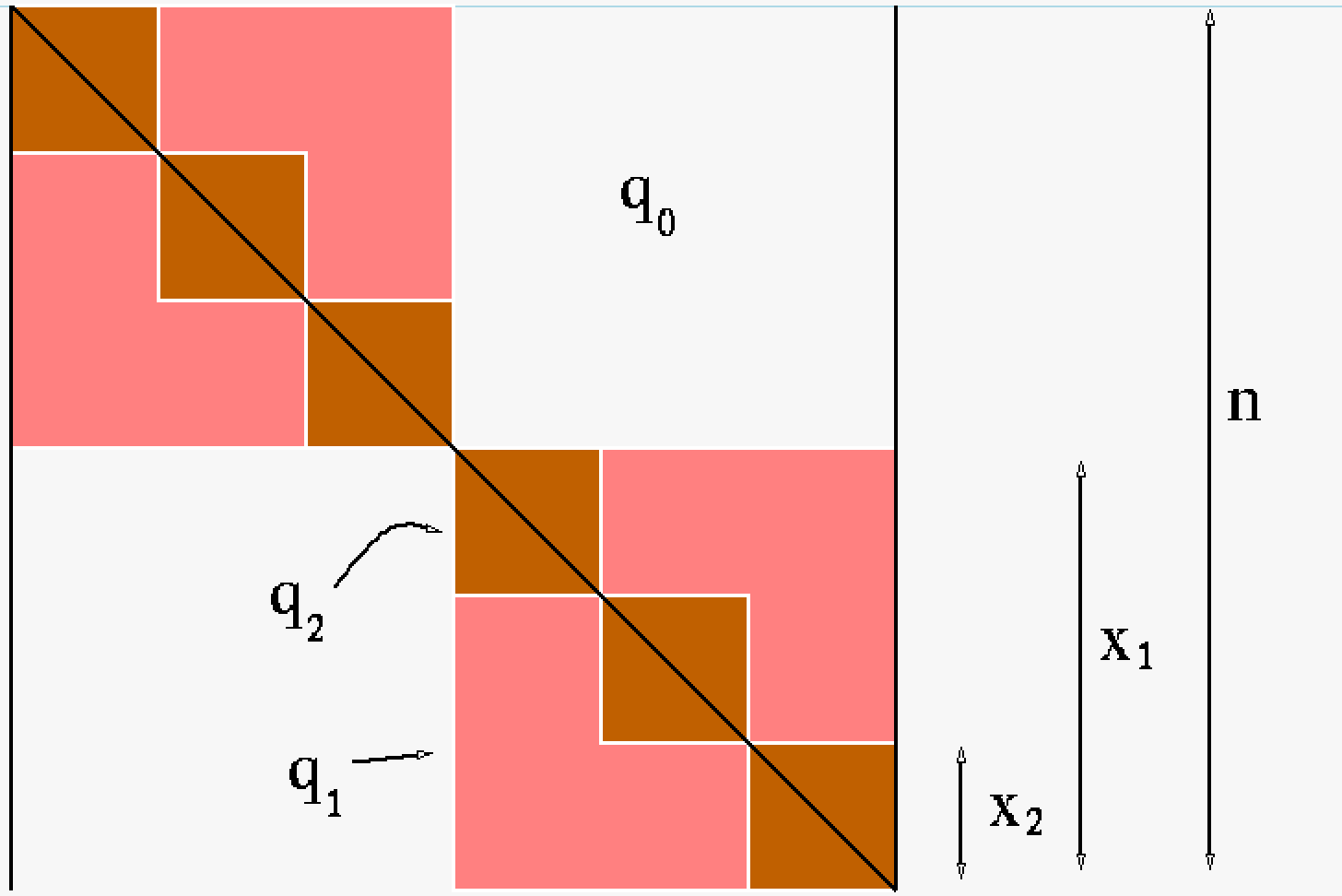}   
    \caption{The famous Parisi hierarchical overlap matrix, encoding the full overlaps' statistics of disordered systems with broken ergodicity.}
    \label{matrix}
    \end{center}
\end{figure}

\begin{figure}
    \begin{center}
\includegraphics[width=8cm]{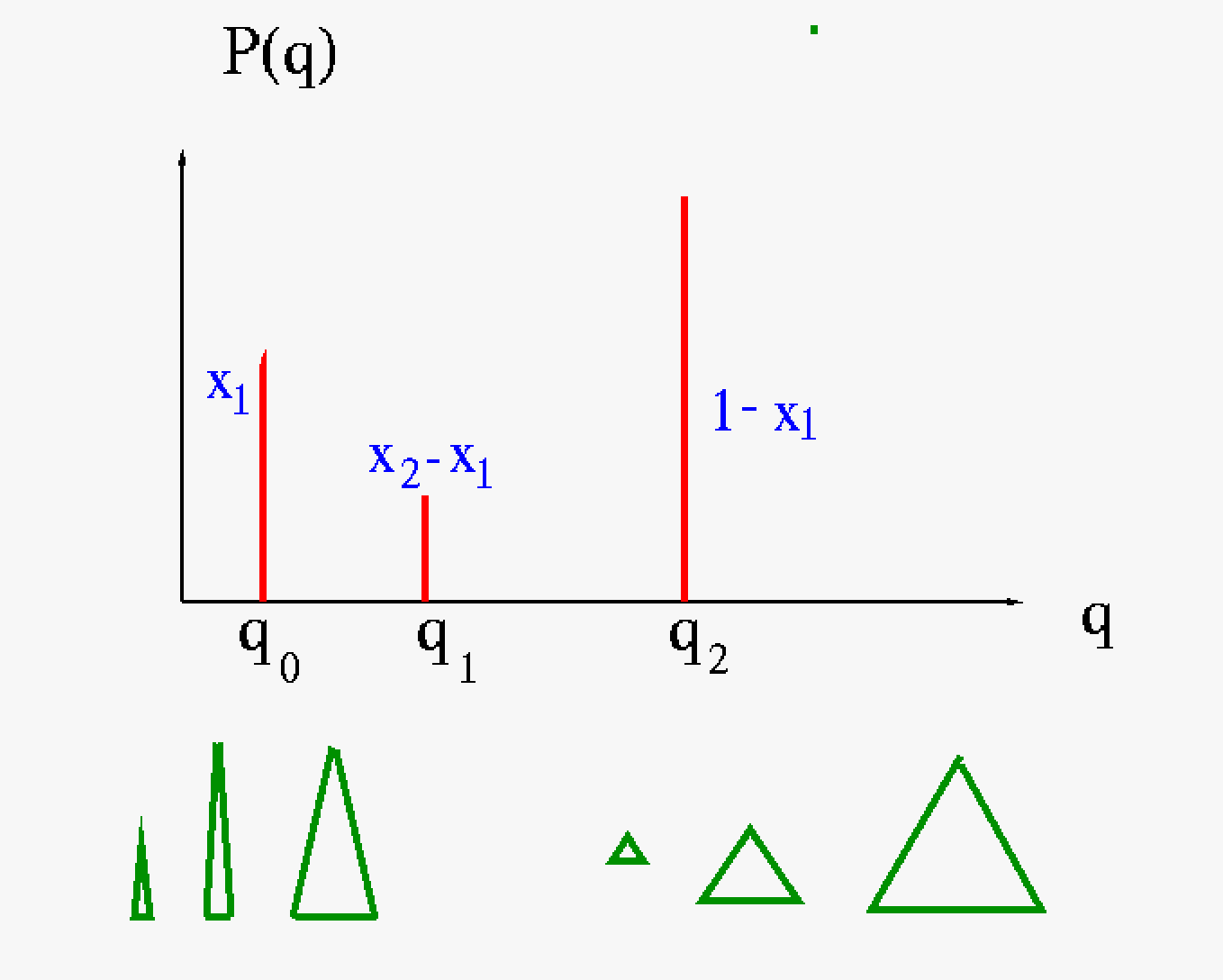}   
    \caption{The possible overlaps and the triangles they determine, according to the ultrametric ansatz.}
    \label{triang}
    \end{center}
\end{figure}

Equations (\ref{GG1},\ref{GG2}) are examples 
of the celebrated Ghirlanda-Guerra identities 
\cite{ghirlanda1998general} that have played a 
crucial role in the mathematical analysis of Spin Glasses. It has been shown by 
Panchenko \cite{panchenko2013parisi} that in their most general form they imply ultrametricity (see Figs \ref{matrix} and\ref{triang})
for any  three states at mutual overlaps $q_{12},q_{23},q_{31}$: 
with probability one in the thermodynamic limit the two smallest overlaps are 
equal (all triangles are isosceles)
  $q_{13}= \min (q_{12};q_{23})$. Moreover, if we consider the conditional probability \cite{MezardParisiVirasoro} we find:
  \begin{eqnarray}
   P(q_{13}|q_{12}=q_{23}=q)&=&\frac{P(q_{13},q_{12}=q,q_{23}=q)}{P(q_{12}=q,q_{23}=q)}\\
   &=&\theta(q_{13}-q)P(q_{13})+\delta(q_{13}-q)x(q).
  \end{eqnarray}
where $x(q)=\int_0^q dq'\; P(q')$ is the cumulative probability. 
Notice that $q_{13}$ is independent of $q$ when it is larger than $q$, and that, as in the case 
of the overlap pairs, the triangle probabilities are fully specified by the function $P(q)$, or equivalently 
by its primitive, $x(q)$. Indeed, this generalizes to arbitrary number of 
replicas: thanks to self-averageness, $x(q)$ provides a sufficient statistics for the overlaps. 

The picture is rather remarkable: in any system where extensive quantities are self-averaging {and the system is stable with respect to small long-range random interactions}, the only 
possibility besides triviality is ergodicity breaking according to the Parisi scheme. 
Originated in an unconventional mathematically analysis of the famous $0\times 0$ replica matrix, the Parisi RSB 
scheme has revealed a deep, and inescapable symmetry of nature, { transcending  the replica trick}.

 \subsection{Overlap locking}
 It can be easily realized that if RSB is present in a system, then the overlap 
  between any two random configurations should be homogeneous in space. 
  This is true in long range systems, where by space we mean the complete graph 
  (SK model) or the Bethe lattice (Viana-Bray model), but also, and especially, 
in any finite dimensional system with short range interaction that could possibly display RSB. 
  Let us specifically refer to this case. Overlap homogeneity means that given two 
  random equilibrium configurations, if one measures the value of the overlap in a 
  given large enough region of space, the same value 
 would be found by an analogous measure in any other region, no matter how far away from 
 the original one. 
 This is a very strong correlation property of equilibrium states. 
 
 Suppose you divide the system in two adjacent parts. 
 The surface interaction term could be disconnected, without any appreciable change 
 in the 
 expectation values of bulk quantities. On the other hand, this term is 
 crucial for the overlap statistics: in absence of the surface interaction
 the overlaps in the two halves are independent variables; in presence of it, the overlaps 
 are locked to take the same values. 
We can generalize this consideration to more general weakly coupled disordered systems. 
The locking of the overlaps to identical values in the 
parts of the same system occurs because of statistical homogeneity of a given sample. 
An analogous locking however also occurs in absence of homogeneity, 
whenever we weakly couple systems with RSB. Suppose we have  
two systems ${\cal S}_1$ ${\cal S}_2$, whose overlap statistics is specified by $x_1(q)$ and $x_2(q)$ 
respectively, not necessarily identical. As before, as long as the systems are not interacting the overlaps are independent. Stability of the RSB state on the other hand requires that the as soon as a weak interaction 
is switched on, the overlaps do lock, and one can define a function $Q_2(q_1)$, which allows to predict the overlap
between two replicas in the system $S_2$, when the overlap is $q_1$ in the system $S_1$. It is clear that 
this relation should be specified by the condition $P_1(q_1)dq_1=P_2(q_2)dq_2$, or equivalently $x_1(q_1)=x_2(Q_2(q_1))$ \cite{franz1992ultrametricity}. The actual functions $P_1(q)$ and $P_2(q)$ on the other hand will be only 
linearly affected by the coupling, and coincide in a fist approximation with the overlap probability distribution in the unperturbed systems. 
{ Here we encounter a first instance of `softness' of the solution: the presence or absence of small, thermodynamically irrelevant 
 terms, leads to a complete rearrangement of the Gibbs weights, without affecting the values of self-averaging quantities. Below we shall find the same property in the context of dynamics.} 
 
\subsection{Marginal stability and Stochastic Stability in Mean-Field}

Within mean-field, two classes of models are known: Models with 'full-RSB', as the SK model,  where the support $Supp(P)$ is a continuous interval, say $[q_0,q_1]$ and models where the support has 'holes'. The most common case 
of this scenario is the case of 'one-step RSB' (1RSB) where only two values of the overlap support the whole probability, this is the case of Potts-glasses and $p$-spin models and for simplicity 
we will refer to it here. We shall address these support values as `skeleton' values of the correlations, and discuss how they play an important role in the correspondence between equilibrium and dynamics.

The physical picture between continuous and 1RSB  cases is very different: in the first case 
the glassy phase is only marginally stable. This can be easily realized: the supports of the 
overlaps distributions are self-averaging \cite{parisi2004distribution}, implying that it is possible to go from a 
given state to any other crossing only sub-extensive barriers  \cite{franz1995recipes,franz1998effective}. Within 
the replica method one finds families of Hessian zero modes associated to small 
variations $P(q)\to P(q)+\delta P(q)$ \cite{temesvari2000reparametrization,de2006random}, a reparametrization transformation that has a deep dynamical analogue as we will see in the following sections. In the second case, conversely, equilibrium states are absolutely stable and are separated by extensive free-energy barriers. The effect of a small but 
extensive random perturbation is very different in the two cases. In the 1RSB cases 
the states are reshuffled: their weight changes but they keep their identity and continue to exist. 
In the full-RSB case the effect is more dramatic, any $O(N)$ perturbation  destabilizes the unperturbed equilibrium states and produce new ones in random positions. A crucial property is that the {\it probability distribution} of the disorder-dependent Gibbs state is only affected in a regular way by the perturbation (i.e. linearly for small perturbations). This regularity property, called Stochastic Stability is very natural and at the same time important, it tells us that the original model is generic, sharing the same physics with a broad class of similar systems. 

\section{Dynamics and Multithermalization}

\subsection{Slow dynamics}

In the dynamic approach we consider a system evolving according to  local reversible dynamics. To 
fix the ideas we can consider the Langevin equation
\begin{equation}
-m_i {\ddot s}_i- \frac{
\partial H({\bf s})}{\partial s_i}   =
 \underbrace{\Gamma_0 {\dot s}_i - \eta_i}_{bath}
\label{motion}
\end{equation}
where $\eta_i$ are uncorrelated Gaussian white noises with variance $2\Gamma_0 T$
and $\Gamma_0$ is the strength of the coupling to the `white'  bath. This is guaranteed to reach eventually equilibrium,  although in the systems that concern us, in times that may diverge with  the system's size $N$. For the considerations that follow, any stochastic reversible dynamics whose update rule just depends on the force $f_i=- \frac{\partial H({\bf s})}{\partial s_i}$ as e.g. Metropolis or Glauber dynamics
would give rise to the same results. 
We shall be concerned with the limit of slow dynamics, which appears, in glassy systems,  in different ways:
\begin{itemize}
\item {\em Aging \cite{Castellani2005}:} Following a quench of the system from a high to a low temperature, at which the equilibration time is infinite,  the system `ages': it evolves slower and slower as the time since the quench elapses.  
The two-time correlations never  become an exclusive function of time-differences.
The large parameter is the smallest `waiting'  time since the quench  $t'=t_w$, that modulates the decay at further time $C(t,t')=C\left(\frac{\tau}{t_w}\right)$: the older (large $t_w$) the system, the
slower the decay. 
In such conditions of long time scales 
local observables depending on configurations at a single time, such as the energy or specific heat, 
magnetization etc.,   only undergo small variations, and in what follows we will consider the ideal situation 
where they can be taken to be essentially equal to their asymptotic values.

\item {\em Driven system \cite{thalmann2001aging,berthier2000two}} When the system is subjected to forces not deriving from a potential -- shear, for example -- it is an experimental fact that aging is interrupted,
in the sense that all functions become time-translational invariant, but slow. 
Their timescale of  decay of correlation then is controlled by the driving rate $\sigma$, the slower
the weaker the drive:     $C(t-t')=C\left({\tau}{\sigma}\right)$
\item {\em Time-dependent disorder \cite{horner1992dynamics}. } Another way to make a system with disorder time-translational invariant is to change the disorder slowly : the small parameter is the timescale $\tau_0$  of change of disorder: $C(t,t')=C\left(\frac{\tau}{\tau_0}\right)$. 
\end{itemize}
In what follows, we refer  as  {\em slow dynamics} to the limit of either long waiting times, small shear strains or slow variation of parameters.
The latter two cases are stationary, all dependence on history is lost,  but quite surprisingly they
are {\em not} equilibrium systems, as we shall see.

In all these systems we consider  correlation $C_{AB}(t,t')$  and linear response function $R_{AB}(t,t')$ (here $t\ge t'$), the average response of $A$ at time $t$ to
a kick of $B$ at time $t'$. 
In the case of spin variables above, correlations and response functions are given by:
\begin{equation}
C_{ij} (t,t') =  \langle s_i(t) s_j(t') \rangle \qquad ; \qquad R_{ij} (t,t') = \frac{\delta}{\delta h_j(t')}\langle  s_i(t) \rangle|_{h=0}
\end{equation}
where $h_i(t)$ is the value at time $t$ of a field conjugated to $S_i$ in the Hamiltonian. 
One in general is interested to the global correlations and response, defined as
$C(t,t') = \frac 1 N \sum_i C_{ii}(t,t') \qquad ; \qquad R(t,t') = \frac 1 N \sum_i R_{ii}(t,t') $.
It is useful to consider the response function in its integral form: 

\begin{equation}
\chi(t,t') =  \int_{t'}^t dt'' \; R(t,t'')
\end{equation} 
representing the response at time $t$ to a field acting on the system from $t'$ to time $t$

 In an equilibrium situation, we have  for all $A,B$ time translation invariance 
 $C_{AB}(t,t')=C_{AB}(t-t')$ and the Fluctuation-Dissipation relation:
\begin{equation}
T R_{AB}(t,t') = \frac{ \partial C_{AB}(t,t')}{\partial t'} 
\end{equation}
Glassy dynamics, even when it is rendered stationary as above, violates these relations. 
In the spirit of the above, it makes sense to define  effective temperatures \cite{CugliandoloKurchanPeliti}  as:
\begin{equation}
T_{AB} (t,t')R_{AB}(t,t') = \frac{ \partial C_{AB}(t,t')}{\partial t'} \qquad \qquad ; \qquad \qquad  T_{AB} (t,t') = \frac{T}{X_{AB} (t,t') }
\end{equation}
In equilibrium  $X=1$ and where $T_{AB}(t,t')=T$, the bath's temperature.
It turns out that these effective temperatures are exactly what a thermometer would measure when tuned to respond to the corresponding observables at the corresponding temperatures \cite{CugliandoloKurchanPeliti}.

\subsection{Reparametrization softness}

Glassy dynamics have a remarkable emergent property that has long been 
recognized: their evolutions may be drastically modified by small perturbations. 
For example, a static perturbation may lead the system to explore 
regions of the configuration space very different 
from the one reached in absence of the perturbations; 
the introduction of a perturbation in a well-aged system leads 
to rejuvenation, modifying the rhythm of relaxation. 
Even more dramatically, as we have mentioned above, the presence 
of a small shear or evolution of disorder
leads to interruption of aging and the system is rendered stationary.

All these are highly nonlinear effects on the system's dynamics. 
Paradoxically, this is does not mean that linear response breaks down: in spite 
a very modified   time-dependence, there are   
families of relations where time is gauged out that are only linearly 
affected by the perturbations.  
When one looks more closely at how this happens, it turns out that this `softness' concerns the {\em pace} of the dynamics, where two point correlations become modified by a {\em time reparametrization} $t\to h(t)$ as follows:
\begin{eqnarray}
C(t,t')  \rightarrow C(h(t),h(t'))\nonumber\\
R(t,t') \rightarrow \dot h(t') R(h(t),h(t'))\nonumber\\
\end{eqnarray}
Or, in general, for a multi-point average $O(t_1,...,) = \langle s_i(t_1) \eta_j(t_2) s_k(t_3) ... \rangle$ a reparametrization acts on each factor as $s_i(t) \rightarrow s_i(h(t))$ and $\eta_i(t) \rightarrow \dot h(t) \eta_i(h(t))$ 

\subsection{Reparametrization-invariant quantities}

\begin{figure}\begin{center}\label{triangle}
\includegraphics[width=7cm]{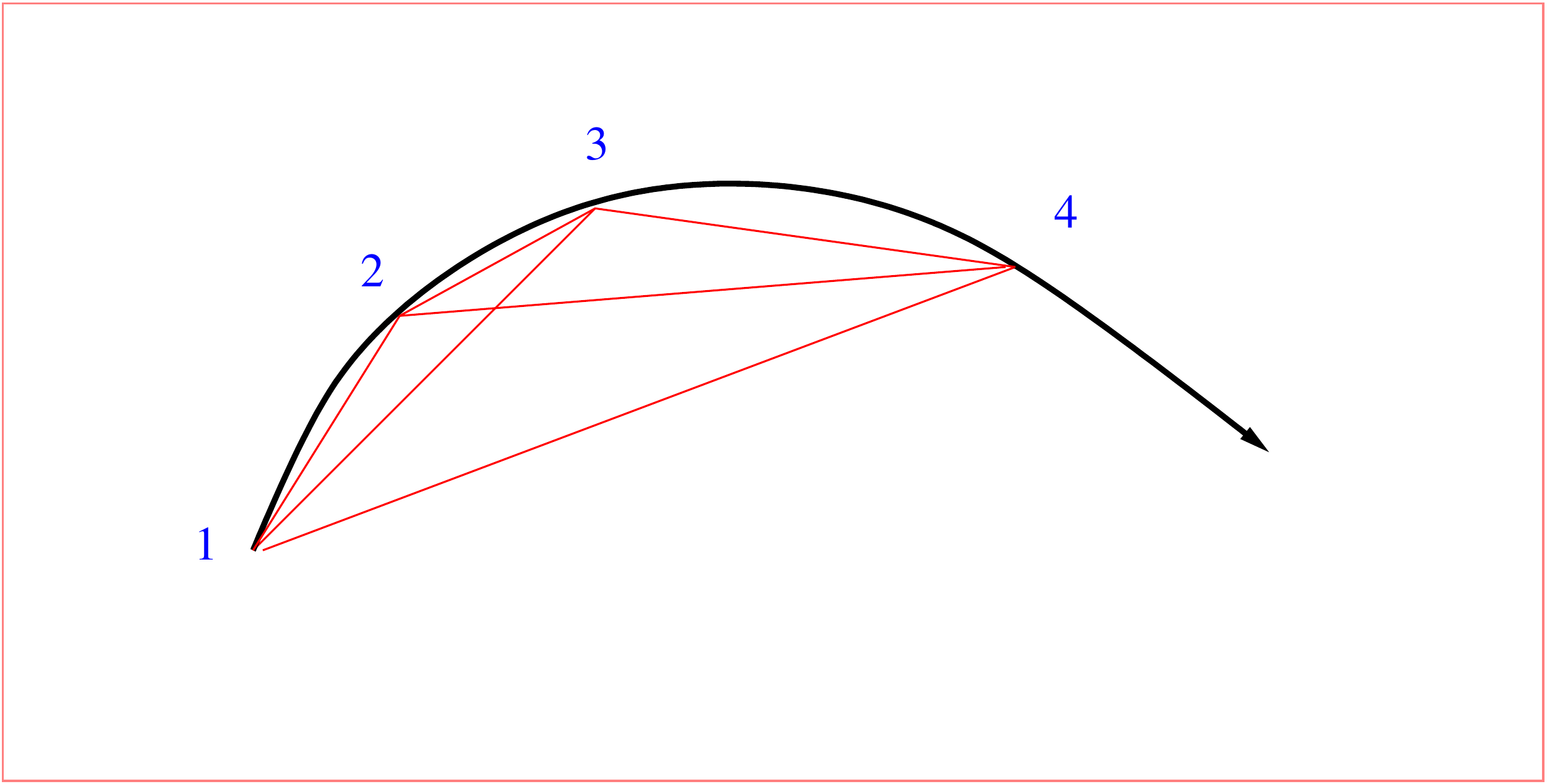}
\caption{The triangle relation $f(a,b)$ is associative: $C_{41}=f[C_{43}, f(C_{32},C_{21})]= f[f(C_{43}, C_{32}),C_{21}]$: the function $f$ is associative.}  \end{center}
\end{figure}

Reparametrization softness means that small perturbations affect, as it were,  
{\em the speed of the film}, but leave the mutual relations `within a scene' invariant. It thus follows that there is a particularly significant sub-ensemble of dynamic  quantities:
those where time is factored out \cite{cugliandolo1994out}, and  thus in the limit of large times they are invariant under reparametrizations $ t \rightarrow h(t)$ ($t\to \infty$), being $h$ any monotonously increasing function of time. 
The first example is rather trivial, but important for the following:  for 
single time quantities $O(t)$, reparametrization invariance implies that they must become independent of time, i.e. to be asymptotically close to their infinite time value.
More significant are the following examples:
\begin{itemize}
\item Given three long, successive  times $t_1<t_2<t_3$, and the corresponding correlations  $C_{21}, C_{32}, C_{31}$, define for large times (see Fig. \ref{triangle}).
$C_{31}  = f(C_{21};C_{32})   = \lim_{t_1 \rightarrow \infty}f(C_{21};C_{32}, t_1)  $, a `triangle relation'.\footnote{More precisely the relation should be thought to be valid in 
the large time limit according to $f(C_{21};C_{32})=\lim_{t_1<t_2<t_3\to\infty \atop C(t_2,t_1)=C_{21}; C(t_3,t_2)=C_{32}} C(t_3,t_1)$ and similarly for the quantities that we discuss in what follows.} 
Similarly for the integrated responses: $\chi(C_{31})  = \tilde f[\chi(C_{21});\chi(C_{32}) ]$, i.e. the  triangle relations $\tilde f$ and $f$ are  isomorphic. 
\item Given any dynamic parameter $X_{AB}(t,t')$  define for large times  $X_{AB}(t,t') \rightarrow X_{AB}\left[C_{AB}(t,t')\right] $. We shall focus on cases
in which this limit is non-trivial $X \neq 1$. 
This also implies that the integrated response becomes a function of the correlation: $\chi(t,t') \rightarrow \chi(C(t,t'))$.
\item Given any two correlations of the system $\bar C(t,t')$ and $C(t,t')$ we write, again in the large-time limit, $\bar C \rightarrow g(C)$ for some $g$.
\end{itemize}
A way of putting this is as follows: a reparametrization-invariant quantity is a function in which times have been
eliminated in favor of a correlation at those times: the particular one used acts as a `clock' with respect to which all other quantities can be referred.

It is very natural to separate those quantities that are reparametrization-invariant from the reparametrizations themselves. A procedure for this may  be implemented  numerically: see \cite{Castillo2002,Castillo2003,Chamon2004,Chamon2002,Chamon2007,Chamon2011}.

\subsection{Time scales}

The reparametrizations we consider are such that they leave the triangle function $C_{13}=f(C_{12},C_{23})$
invariant. The function $f$ allows us to define, in a natural way, time scales in a reparametrization-invariant way. \\

The function $f(a,b)$ may be easily seen to be associative (see Fig \ref{triangle}).
It is possible to classify all associative functions, it turns out that the correlation has special `skeleton' values, that for simplicity we write here as a 
discrete and growing series, $q_1,...,q_k$ that divide the possible correlation 
in intervals $(q_1,q_2),(q_2,q_3)...(q_{k-1},q_k) $ with the property that 
if $C_{12}$ and $C_{23}$ belong to two different intervals then $f(C_{12},C_{23})= \min(C_{12},C_{23})$.  
If, on the contrary, $a$ and $b$ are 
in the same interval, another law applies. 
Each interval corresponds to a time scale, defined in a reparametrization-invariant way. 
{\em The correspondence between Parisi and a dynamic, `multi-thermalization' scheme (to be defined below)  holds at the level of these `skeleton' values. }  
Let us see some examples of two, three and infinitely many scales where the set of skeleton 
values corresopndo to a continuous interval.

\paragraph{Two scales}~\\

This is the most usual case. An example is when the correlation    $0<C\le 1$ and there is a value $q$ such that for the interval $q\le C\le1$
the correlation is much faster than for the interval $0\le C <q$. 
We have, for example: 
\begin{itemize}
\item For a stationary case $C(t-t') = (1-q)\; \bar A(t-t') + q \; \bar B\left(\frac{t-t'}{H(\tau_0)}\right)$, where $H$ is a growing function of $\tau_0$.
\item  For an aging case $t>t'$:\\ $C(t,t') = (1-q)\; A(t-t') + q \; B\left(\frac{L(t')}{L(t)}\right) = (1-q)\; A(t-t') + q \; B\left(e^{h(t)-h(t')}\right)$   \qquad \qquad 
\end{itemize}
where $(A,B,\bar A , \bar B)$ are functions decreasing from  one to zero as their argument goes from zero to infinity. The time reparametrization  $h(t) = \ln L(t)$ brings the aging form into a time-translational invariant form.

\paragraph{Three scales}~\\

Again, the correlation    $0<C\le 1$ and there two values $q_0$ and $q_1$  such that for the interval $q_1\le C\le1$
the correlation is much faster than for the interval $q_0\le C <q_1$, itself much faster than  $0 \le C <q_0$
We have, for example, for the stationary state: 
\begin{itemize}
\item $C(t-t') = (1-q_1)\; \bar A(t-t') + (q_1-q_0) \; \bar B\left(\frac{t-t'}{H(\tau_0)}\right)+q_0 \; \bar{\bar B}\left(\frac{t-t'}{\bar H({\tau_0})}\right) $
\end{itemize}
where $ \bar{\bar B}$ is also decreasing from  one to zero as their argument goes from zero to infinity. The timescales are nested as $\tau_0 \rightarrow \infty$: $\bar H( \tau_0 )\gg H(\tau_0)\gg 1$.

The function $f$  is the function $\min$ for correlations
in any two different scales. \\

\paragraph{A continuum of scales}~\\

 An important case is when there is a dense set of values of correlation in which for all values of correlation 
\begin{equation} f(C_{21},C_{32}) = \min [ g(C_{21}),g(C_{32})] \label{uuu} \end{equation} 
holds. An example is:
\begin{equation}
C(t,t')= {\cal{C}}\left(\frac{\ln(t-t' +t_0)}{\ln \tau_0}\right)
\label{log}
\end{equation}
where $t_0$ is a constant. This form satisfies (\ref{uuu}) when $\tau_0 \rightarrow \infty$ \footnote{Note however that, confusingly,  $C(t,t')= {\cal{C}}\left(\frac{\ln t'}{\ln t}\right)$ is only {\em one} scale!}.
Note how simple ultrametricity appears in dynamics.
The Sherrington-Kirkpatrick model (stationarized by shear or by parameter evolution, see next subsection) follows this law with astonishing precision.
\cite{contucci2020stationarization}.

\subsection{Bringing a system to a stationary form}

Shear and evolving disorder render a system stationary. It came as a surprise that 
for small drive the resulting stationary system is
by no means close to an equilibrium one (as in usual systems with small currents)  but 
rather a time-reparametrized version of the aging system.

For many technical purposes it is convenient then, at least in numerical computations, to {\em stationarize'} the system. The shear strength or the speed of evolution of the disorder becomes then  the control parameter, substituting the self-generated waiting time of the aging situation.

\subsection{The Multithermalization scheme}

Having defined timescales and effective temperatures, we are ready to state the multithermalization scheme. For all observables $A,B$ the long-time response and correlation functions define an effective temperature $A,B$ as $T_{AB}(t,t')$ which
we may write using a correlation as a `clock' $T_{AB}(t,t') \rightarrow T_{AB}(C_{AB})$.\\

$\bullet$ Timescales defined on the basis of the correlation of any pair of observables are the same.\\

$\bullet$ The effective temperatures are, within a timescale, the same for all pairs of observables.

\section{Susceptibility Lego}

In this section we  discuss naive linear response theory (LRT), 
and show that both in statics and in dynamics, the non-ergodic order parameter, 
 $x(q)$ and $X(q)$ can be 
viewed as the generating functions, respectively in equilibrium and in 
dynamics of a set of responses to random 
perturbation to the Hamiltonian.  Let us add a to the Hamiltonian a perturbation 
of the form of a $p$-spin long range Hamiltonian
\begin{eqnarray}
&&H \rightarrow H+\epsilon_p H_p\\  
&&H_p=-\frac{1}{N^{(p-1)/2}} \sum_{i_1,...,i_p} h_{i_1,...,i_p}  S_{i_1}...S_{i_p} 
\end{eqnarray}
The perturbations $H_p$ are bulk self-averaging quantities, 
and we have seen that in equilibrium take values 
$I^{(p)}_{eq}=\mathbb{E}(\omega( H_p)/N=\beta \epsilon_p (1-\int dq \; P_\epsilon(q) q^p)$. 

As usual in linear response theory we can consider the limit of small $\epsilon$,  $
\chi^{(p)}_{eq}=\lim_{\epsilon_p\to 0} I^{(p)}/\epsilon_p=T  \lim_{\epsilon_p\to 0}\lim_{N\to\infty} 
\frac{1}{N} \frac{\partial^2}{\partial\epsilon_p^2}\log Z$. 
If this limit is regular, simple measures of the the energy absorbed by the 
perturbation inform us on the correlations in the unperturbed system. 

Analogously, in dynamics, we can write for the finite time expected value of $H_p$
\begin{eqnarray}
 I^{(p)}(t)&=& \mathbb{E}\langle H_p(t)\rangle_{dyn}\\
&=& \epsilon_p \int_0^t ds\; R(t,s) p C(t,s)^{p-1}
\end{eqnarray}
and introducing $\beta X(q)=\lim_{t,s\rightarrow\infty\atop C(t,s)=q} 
\frac{R(t,s)}{\partial C(t,s)/\partial s}$ we have
\begin{eqnarray}
 \lim_{t\to\infty}I^{(p)}(t)&=& \beta \epsilon_p \int dq \; X(q) \; p q^{p-1}
\end{eqnarray}
As in the case of equilibrium $X$ is the generator function of response 
{\it in presence of the perturbation} and if its $\epsilon\to 0$ limit is regular, 
we have information about the response in the unperturbed system. 

\subsection{Triangles}
One can define higher order susceptibilities that in equilibrium allow one to reconstruct 
the probability distribution of several 
replicas, and in dynamics reflect the relations of correlations at different times.  
Particularly interesting are the ones associated to the probability of triangles $P(q_{12},q_{13},q_{23})$, which within Parisi RSB is concentrated on ultrametric triangles. 
In FMPP the triangle probabilities were reconstructed by considering average values of mixed combinations of the two different $H_p$, where the tensors defining the interactions are contracted 
over a subset of indexes, namely, in presence of the perturbations $H_{n+m}$ and $H_{n+l}$ in the Hamiltonian one considers the average value of the observable 
\begin{eqnarray}
O_{n;m,l}= \sum_{i_1,...,i_{n+m+l}}J^{(n+m)}_{i_1,...i_n,j_{n+1},...,j_{n+m}}
 J^{(n+l)}_{i_1,...i_n,k_{n+1},...,k_{n+l}}S_{j_{n+1}}...S_{j_{n+m}}S_{k_{n+1}}...S_{k_{n+l}}. 
\end{eqnarray}
where $n$ among the indexes of the two tensors, of rank $n+m$ and $n+l$ respectively, are contracted between them, and the others with the spins of the system. 

Straightforward integration by part allows to see that in equilibrium 
\begin{eqnarray}
\frac 1 N \mathbb{E} \langle O_{n;m,l}\rangle=\beta^2 \epsilon_{n+m}\epsilon_{n+l}\left(
1-\langle q^{n+m}+q^{m+l}+q^{n+l}\rangle+2\langle q_{12}^n q_{13}^m q_{23}^l \rangle
\right)
\end{eqnarray}
while in dynamics 
\begin{eqnarray}
\frac 1 N \mathbb{E} \langle \delta O_{n,m,l}(t)\rangle=
\epsilon_{n+m}\epsilon_{n+l} \int_0^t dv \int_0^t du &&([lC(t, v)^{l-1} R(t, v)][C(v, u)^m ][nC(t, u)^{n - 1}
R(t, u)] +\nonumber\\
&&[lC(t, v)^{l-1} R(t, v)][mC(v, u)^{m-1} R(v, u)][C(t, u)n]
+\nonumber\\
&&[C(t, v)l][mC(u, v)^{m-1} R(u, v)][nC(t, u)^{n-1} R(t, u)])
\end{eqnarray}
Since $\langle O_{n;m,l}\rangle$ is not directly a response function 
(i.e. a derivative of the free-energy) its equilibration was considered 
plausible, but the relation between triangles was not-completely proven within LRT. 
To remedy to this weakness, in \cite{kurchan2021time} it was shown that in fact the triangular correlations could also be generated as responses to perturbations: 
it suffices to perturb the Hamiltonian with the following term
\begin{eqnarray}
  \delta H_{n,m,l}=(\gamma_{ml} O_{n;m,l}+\gamma_{nl} O_{m;n,l}+\gamma_{nm} O_{l;n,m})
\end{eqnarray}
that generates the triangles at the cubic order in the $\gamma$'s. \\





\paragraph{2. Overlap equivalence between sublattices}~\\   

Another generating function that has proven  useful is the following
Let us first consider a lattice system, which we divide in four sublattices. 
whose components we shall call $s^{(1)}_i$, $s^{(2)}_i$ and $s^{(3)}_i$ and  $s^{(4)}_i$. Adding to the energy a term
\begin{eqnarray}
S&=&   \frac{\gamma}{N}  \sum_{ij} \left(  h^{(1)}_{ij}  h^{(2)}_{jk}   s^{(1)}_k s^{(2)}_i + h^{(2)}_{ij}    h^{(3)}_{jk} s^{(2)}_i s^{(3)}_k+  h^{(3)}_{ij}   {h}^{(4)}_{jk}s^{(3)}_i s^{(4)}_k\right)
\nonumber\\
 \end{eqnarray}
 with the $h^{(\ell)}_{ij}$ random Gaussian variables and  computing the following correlation and its associated response:
\begin{equation}
C^{(4)}(t,t') =\frac{1}{N^2} \sum h^{(1)}_{ij}    h^{(4)}_{jk} \left\langle s^{(1)}_i(t) s^{(3)}_k(t')  \right\rangle_\gamma \qquad  
\end{equation}
This generating function may be generalized to any number of sublattices, and it is useful
to show that there is overlap equivalence if and only if the timescales at different effective temperatures are widely separated.

\section{The nucleation argument}

As we have stressed, in the large limit,  even in presence of ageing, 
the values of  extensive and local single time observables should tend to asymptotic values. 
In mean-field systems with long range interactions
these may or may not coincide  with the corresponding values in equilibrium. 
In systems with finite range interactions in finite dimensions, on the other hand, 
these values should coincide with their Boltzmann-Gibbs equilibrium values. 
The usual argument which is advanced is that in finite dimensions there are always 
local relaxation mechanisms allowing for equilibration. As a consequence, for the bulk  free-energy, 
 the infinite-time limit and the thermodynamic limits commute. 
The same should be generically true for its derivatives, such as  internal energy and static responses. 
Notice that this argument tells that any possible broken ergodicity state should be only marginally stable,  since the introduction of a small extensive random perturbation destabilizes the old equilibrium states 
to the advantage of new ones with a more favorable energetic balance between old and  new  energy. 

This does not of course prevent the practical existence of systems such as structural 
glasses where the relaxation times at low temperature are so high that the  energy density 
remains off-equilibrium on geological scales. Notwithstanding this, the limiting 
situation of aging close to equilibrium is a useful conceptualization to understand how glasses explore the configuration space. Whether in actual experimental systems on lab 
time scale one is close 
or not to this situation has to be decided case by case, depending on the system the preparation and the external parameters. We believe that this should be the case 
for example of spin glasses below, but close enough to the critical temperature. In these 
systems local equilibrium develops on growing scales $\xi(t)$ and the asymptotic 
situation may be observed if the growth of this scale is not too slow.
We have seen that a possible way to measure $P(q)$ function consists in introducing 
{\bf long range} random perturbations. If one admits that these perturbations indeed 
equilibrate (at the level $\epsilon$) in off equilibrium dynamics, then one could 
reconstruct the Parisi function from simple off-equilibrium measures of the response.  
Unfortunately, the nucleation argument does not hold in presence of long range interaction, and one could cast doubts that small long range perturbations would induce metastable states with free-energy density $O(\epsilon)$ higher than the equilibrium one. These could trap the off-equilibrium dynamics, thus spoiling the static dynamic equivalence. Thinking about what it could go wrong, this appears as a very remote possibility for small $\epsilon$. 
Within mean-field, with a long range unperturbed Hamiltonian, where inducing metastability is certainly easier that in finite D, we do not know any case where the introduction of a perturbation induces a difference of order $\epsilon$ between the value of the asymptotic dynamical energy and its equilibrium counterpart.

To obviate to this problem however, one can look for local perturbations giving rise to the same 
correlation functions within Linear Response. In FMPP it was suggested that this could be achieved 
diluting the $p$-spin perturbations so that they reduce to finite range interactions. 

Let see how this works for $p=2$, the generalization to arbitrary $p$ being straightforward (and discussed in detail in FMPP).
To fix the idea consider a spin system on a $D$ dimensional square lattice of size $L$,  $\Lambda=\{1,...,L\}^D$. One can divide the lattice in two halves, $\Lambda_l$ and $\Lambda_r$, where the one of the coordinates, say in direction of the x-axes $e$, take the values $\{1,...,L/2\}$ and ${L+1,...,L}$ respectively.    
Let us add to the original Hamiltonian a perturbation of the form $\epsilon_2 H_2$, with
with 
\begin{eqnarray}
 H_2=\sum_{x\in \Lambda_-}h_x S_x S_{T(x)} 
\end{eqnarray}
where the $h_x$ are independent Gaussian random variables with unit variance, and $T(x)$
is a translation of $x$ of length $L/2$ in direction $e$. The thermal expectation 
value of the perturbation $H_2$ gives a contribution to the internal energy of the system which is extensive and self-averaging, i.e., independent (in the thermodynamical limit) of the particular realization of the disorder contained in either $H$ or $H_2$. The interaction $H_2$, which looks long range, is, in fact, a local perturbation in a different space.
Let us rename the spins in the right-hand part so that if $x\in \Lambda_l$ then
$T(x)\in \Lambda_r$  and $S_{T(x)}=S'_x$. The total Hamiltonian can now be written as
\begin{eqnarray}
\label{cinque}
 H[S,S']=H_l[S]+H_r[S']+B[S,S']+\epsilon_2 \sum_{x\in \Lambda_-}h_x S_x S'_x
\end{eqnarray}
The Hamiltonian $H_l$ and $H_r$ refer, respectively, to the
spins in $\Lambda_l$ and $\Lambda_r$. The term $B[S,S']$ is a surface term
whose presence does not affect the average of $H_2$. Dropping it, the Hamiltonian (\ref{cinque}) characterizes a spin system of size $L/2$ with two spins $S_x$ and $S_x'$ on each site, and a purely local interaction. We have traded the long range interactions with a local coupling between among non-interacting or weakly coupled systems with different disorder and equal number of spins. We would like now to show that within LRT $H_2$ continues to be associated to the second moment of the overlap PDF. Simple integration by part in equilibrium reveals that 
\begin{eqnarray}
 \langle H_2/N \rangle =\beta \epsilon_2 \frac 1 N \sum_{x\in \Lambda_l} \left[1-\mathbb{E}\left(\langle S_x S_{T(x)}\rangle^2_\epsilon\right)\right]. 
\end{eqnarray}
In FMPP it was argued that the limit for 
$\epsilon\to 0$ of 
$\mathbb{E}\left(\langle S_x S_{T(x)}\rangle^2_\epsilon\right)$ 
{\it both in presence or in absence of the boundary term $B[S,S']$ }should be equal to the 
value that this correlation function takes for strictly zero coupling $\epsilon=0$ in presence of $B[S,S']$. The small coupling between the systems must have the same locking effect on the overlap as the surface interaction that we have discussed in the introduction.

\subsection{Overlap equivalence if and only if timescale-separation}

The property of overlap equivalence is very natural in the Parisi scheme. One has a distribution of states with their Gibbs weights.
One now assumes that two pairs of states $(a,b),(c,d)$ having the same mutual  overlap measured with one observable $q_{AA}(a,b)=q_{AA}(c,d)=q$,  will have also the same mutual overlap
measured with any other observable  $q_{BB}(a,b)=q_{BB}(c,d)=\bar q$. 
This means that there is a universal function that
gives, for every state,  the second overlap if one knows the first:
\begin{equation}
    \bar q = g (q)
\end{equation}
Clearly, we have:
\begin{equation}
   \bar P(\bar q) d \bar q = P( g (q)) dq  \qquad \rightarrow \qquad   \bar P(\bar q) = P(g(q)) \frac{1}{g'(q)}
\end{equation}
We may rewrite this putting $\frac{dx}{dq}=P(q)$  and $\frac{d \bar x}{d \bar q}=\bar P(\bar q)$ as:
\begin{equation}
x(q)= \bar x(\bar q) \label{oveq}
\end{equation}

The property (\ref{oveq}) has the dynamic counterpart:
\begin{equation}
X(C)= \bar X(\bar C)  \qquad \rightarrow  \qquad \bar T(\bar C(t,t')) = T (C(t,t'))\label{oveq1}
\end{equation}
i.e. the effective temperature obtained from the fluctuation-dissipation relation of two observables at given times coincide.

This is very striking: a very similar relation arises from two apparently different arguments. And yet, using the second of the generating functions above,
we find that there is a complete correspondence between $g, X, \bar X$ in the dynamic and in the equilibrium context.

In the dynamic case, we may also prove a strong result \cite{kurchan2021time}: overlap equivalence holds {\em if and only if} there is a wide separation of timescales
associated to two different temperatures for each observable. The proof is not complicated: suppose we have two observables that have the same $X(C)=\bar X(\bar C)$ at all times,
but the times for different $X$'s are not really separate. Then, one can show that by coupling the observables one may construct a new correlation $\bar{\bar C}$ that
does not have $\bar{\bar X}$ at the same times. In other words, {\em Multithermalization is only consistent if the timescales of different temperatures are widely separated}.

\section{Commuting limits
}

We have seen that within naive perturbation theory, under the hypothesis of equilibration of 
bulk quantities, dynamics and equilibrium are in strong correspondence and the study of the 
dynamical response can be used to reconstruct the equilibrium Parisi function, which is an average over disorder realizations. 
This is at first sight strange, one has a relation between 
conceptually different kinds of objects that are relevant in different
time regimes (in and out of equilibrium, respectively), when the system explores 
very different regions of phase-space. 

In usual applications of LRT thermodynamic states are supposed to be stable against the introduction of a perturbation. For example if we introduce 
a small positive magnetic field $h$ in the low temperature Ising model we have 
\begin{eqnarray}
 \langle S_i\rangle_h=\langle S_i\rangle_+
 +h\beta\sum_j (\langle S_i S_j\rangle_+ -\langle S_i \rangle_+\langle S_j\rangle_+)
\end{eqnarray}
Typical configurations of the perturbed systems are close to the unperturbed ones in the phase chosen by the field. 

In disordered system by contrast, when a random perturbation is introduced this does not hold, the Gibbs state is marginally stable and 'chaotic' \cite{fisher1986ordered,kondor1989chaos}: 
the perturbation implies a complete reorganization of the Gibbs measure, which concentrates on regions with zero overlap with the unperturbed states 
\begin{eqnarray}
 \frac 1 N \sum_{i=1}^N \mathbb{E}(\langle S_i\rangle\langle S_i\rangle_0)=0. 
\end{eqnarray}
Yet these regions should correspond to close values of the bulk quantities and similar statistical properties. The marginal stability of the Gibbs 
state should then correspond to the stochastic stability of the metastate and $P_\epsilon(q)\to P(q)$. In dynamics the situation is similar: the chaotic 
property imply that no matter how small a perturbation, a perturbed trajectory evolving with the same thermal noise as an unperturbed one, 
will stay close for a certain time, but will eventually diverge when time in large enough. Moreover, any random perturbation introduced after evolution 
up to a waiting time $t_w$ leads to a free-energy increase and hence to a rejuvenation of the system. This means that if we consider a field acting 
from $t_w$ to $t$, the limit for $h\to 0$ of the magnetization at time $t$, 
$\chi(t,t_w)=\lim_{h\to 0} m(t,t_w,h)/h$ is not uniform in time.  As in the case of statics this does not prevent to use 
LRT. We just have to consider reparameterization invariant quantities (e.g. $\chi_{dyn}(q)=\int_q^1 dq\; X_\epsilon(q)$), whose limit $\epsilon\to 0$ should be 
regular as the perturbation tends to zero.

\section{The whole Parisi $\leftrightarrow$ multithermalization scheme needs time reparametrization invariance}

All reparametrization-invariant relations between macroscopic quantities should coincide in dynamics and statics, where in statics only the `skeleton' values are present.
Ultrametricity, overlap equivalence and $x(q)$ are three examples.
As we have mentioned above, the reciprocal is  also true: if there is no timescale separation between {\em different} temperatures , there is no multithermalization, and no equilibrium overlap equivalence.  
There is however a problem.
Consider two  systems brought into weak contact from the beginning, for example two different  lattice models, coupled locally so as to obtain a single model with two sublattices. Assume that at the same times the separate systems have non-coincident temperatures. This is perfectly possible, since the systems are independent.
Now, couple weakly the two systems: should we conclude that  the coupled system, for which there is no distinction between observables of one or the other system,  
violates the multithermalization scenario - and, a fortiori, the Parisi scheme?  If this were so,  both would be fragile to the point of irrelevance.
The answer is surprising: the timescales of the systems rearrange so that different  temperatures happen at different scales: thus, the combined system conforms to the scenario. 
But this needs to happen even  {\em for infinitesimal coupling}, and it can only be possible if the system is  `soft' with respect to time-rearrangements of each temperature separately: in other words, it has to have   independent time-reparametrization invariances  in the slow dynamics limit. Such invariances, which we have already described above,  where first reported 
by Sompolinsky and Zippelius \cite{sompolinsky1982relaxational,sompolinsky1981dynamic} some forty years ago, and have recently had a crucial role in the interpretation of the SYK model \cite{SachdevYe1993} as a toy model of holography 
\cite{KitaevLectures,Maldacena2016SYK}.

\begin{figure}
    \begin{center}
   \includegraphics[width=6cm]{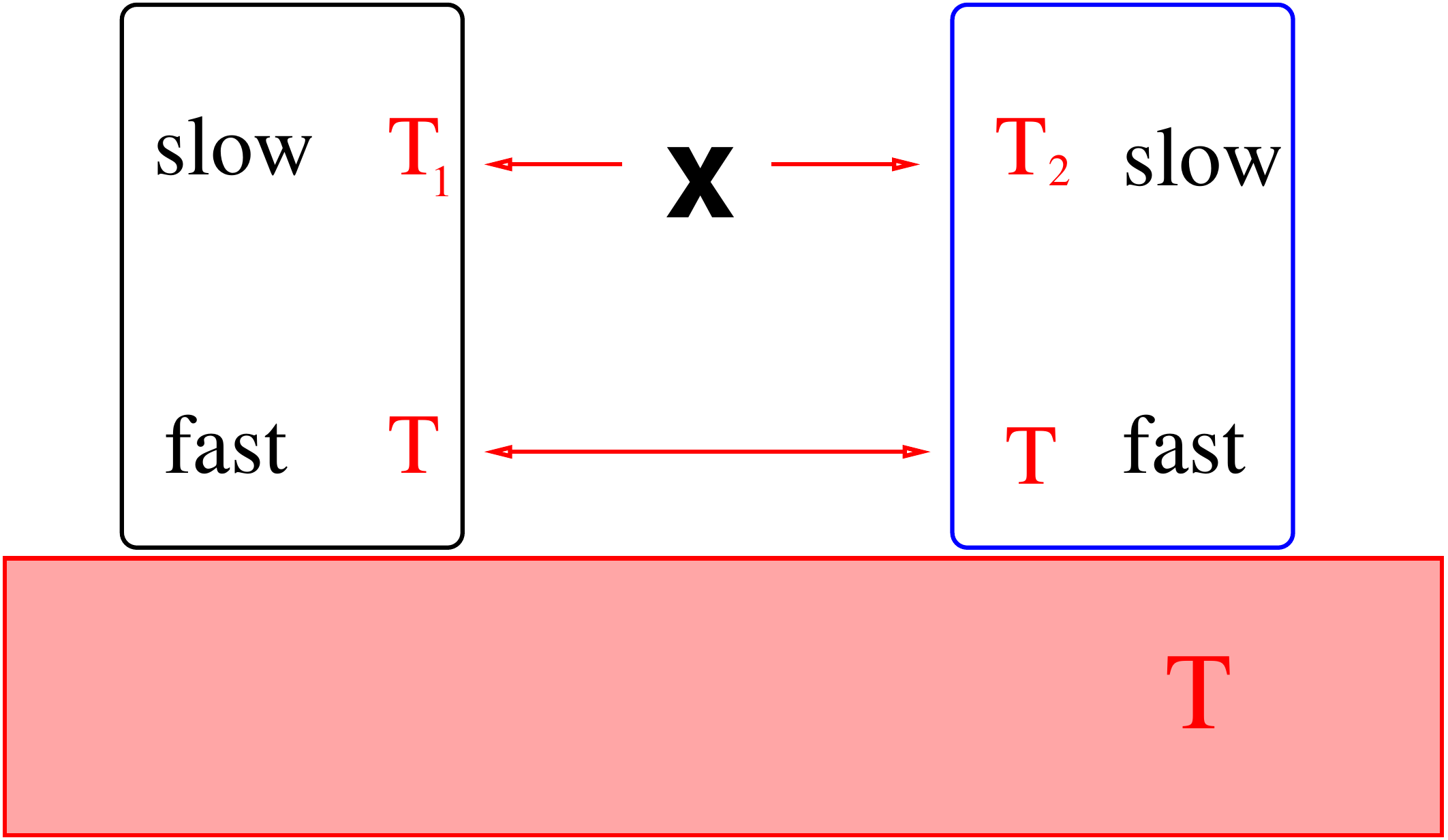} \hspace{1cm}\includegraphics[width=6cm]{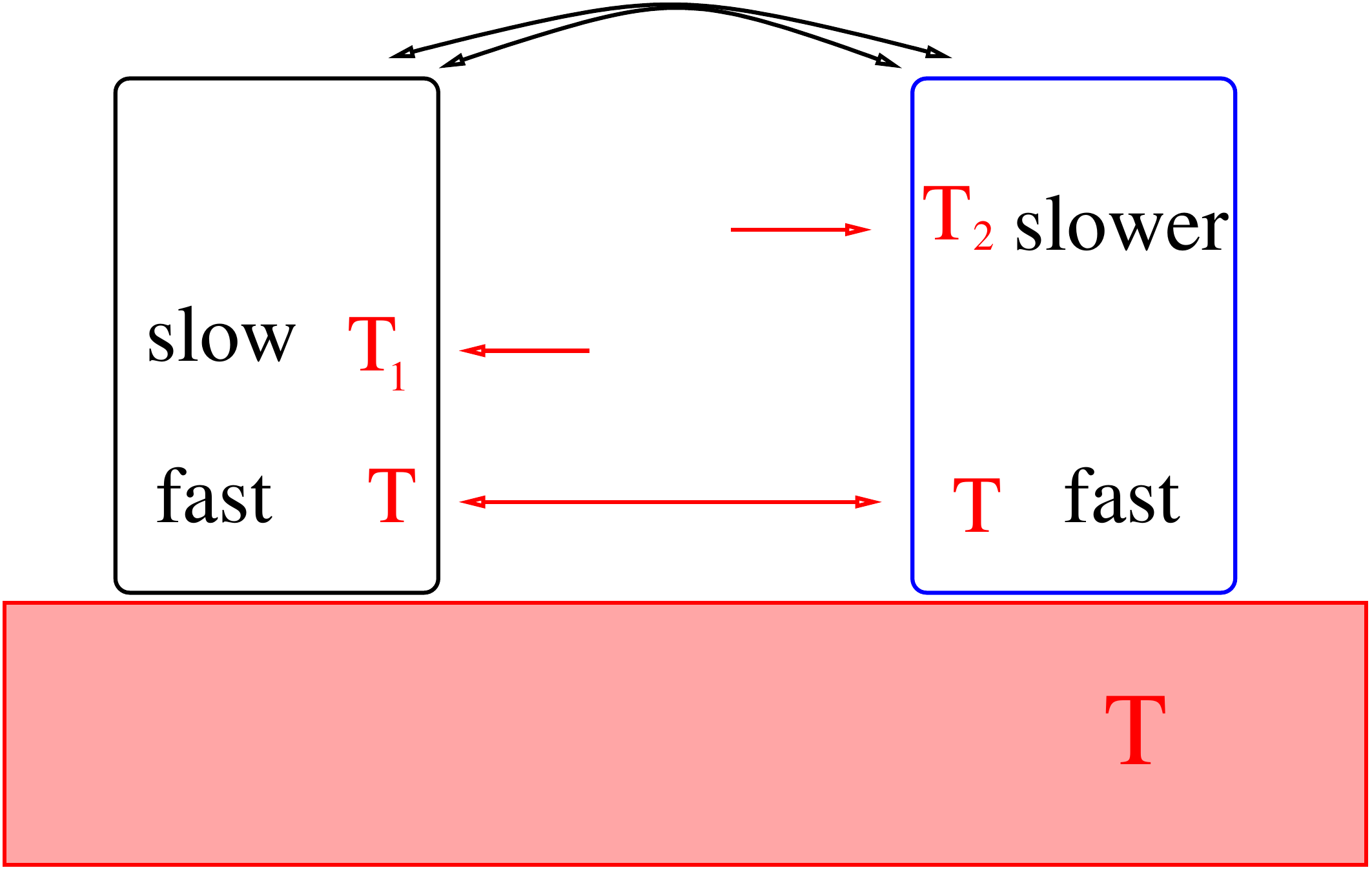}
    \caption{Caption}
    \label{fig:my_label}
    \end{center}
\end{figure}

In conclusion, we realize that reparametrization softness is crucial for the consistency of the multi-thermalization scenario. But we have seen here that this scenario is in a one-to-one correspondence with
the Parisi construction with replicas, at least in finite dimensions. We are thus led to ask ourselves, how do these reparametrizations appear in a replica treatment, where there is no time?   To this we turn in the next section.

\section{Reparametrization invariance and replica space}
     \subsection{A consequence of ultrametricity}

In aging experiment the actual perturbation allowing to measure the function $X(q)$ is just an applied magnetic field. A small magnetic field $h$ acting from time $t_w$ to time $t$
induces a magnetization at time $t$ which is 
\begin{eqnarray}
    m(t,t_w)/h= \chi^+(t,t_w)=\int_{t_w}^t ds\; R(t,s){\rightarrow\atop t,t_w\to\infty; C(t,t_w)=q}\int_q^1 dq' X(q').
\end{eqnarray}
    
    This suggests in equilibrium to consider the response to a magnetic field of a system $S_1$ which is kept to a fixed overlap $q\in {\mbox {Supp}}(P)$ with a
    configuration $S_0$ which is well equilibrated {\it in absence} of the perturbation.

    Namely, considering an external random field term $\delta H=-\epsilon_1 \sum_i h_i S_i$ with $\mathbb{E}h^2=1$
    we would like to compute
    \begin{eqnarray}
     \chi^+(q)=\frac{1}{\epsilon_1} \mathbb{E}
     \sum_{S_0}\frac{e^{-\beta H{[S_0]}}}{Z} 
     \sum_S 
     \frac{
     e^{-\beta (H{[S]}+\delta H[S])}
     \delta(Q(S,S_0)-q)\sum_i h_i S_i} {Z_1[S_0]} 
    \end{eqnarray}
    It is simple to see that within Linear Response Theory the response this is given by 
    \begin{eqnarray}
     \chi^+(q)=\beta\left(1-\mathbb{E}\;
     \left\langle
     \frac{
     \langle
     q_{12} \mathbbm{1}_{q_{10}=q_{20}=q}
     \rangle_{12}
     }{[\langle
         \mathbbm{1}_{q_{10}=q}
         \rangle
     ]^2}
     \right\rangle_0
          \right)
    \end{eqnarray}
where we have denoted as $\langle\cdot\rangle_{a,...}\cdot)$ the Boltzmann average with respect to the replicas $a,...$       Remarkably, using the Ghirlanda-Guerra identities, it is possible to show that 
    this last quantity is equal to 
    \begin{eqnarray}
    \mathbb{E}\;\left\langle
    \frac{
     \langle
     q_{12} \mathbbm{1}_{q_{10}=q_{20}=q}
     \rangle_{12}
     }{[\langle
         \mathbbm{1}_{q_{10}=q}
         \rangle
     ]^2}
     \right\rangle_0
    = 
        \frac{\mathbb{E}\left(
     {\langle q_{12} \mathbbm{1}_{q_{10}=q_{20}=q}\rangle_{012}}\right)}
     {\mathbb{E} \; \langle
     {[
     \langle
     \mathbbm{1}_{q_{10}=q}
     \rangle_1
     ]^2}
     \rangle_0
     } =\int dq_{12}\; P(q_{12}|q_{10}=q_{20}=q) q_{12}
    \end{eqnarray}
    
The conditional probability
$P(q_{12}|q_{10}=q_{20}=q)$ is fixed by ultrametricity 
$P(q_{12}|q_{10}=q_{20}=q)=x(q) \delta(q_{12}-q)+\theta(q_{12}-q)P(q_{12})$. 
Which in turn implies that  
    \begin{eqnarray}
     \chi^+(q)=\beta \int_q^1 dq\; x(q). 
    \end{eqnarray}
This has to be compared with the dynamical susceptibility at time $t$ to a field acting from time
$t_w$ to $t$, in the long time limit having fixed $C(t,t_w)=q$, namely
   \begin{eqnarray}
     \chi_{dyn}^+(q)=\beta \int_q^1 dq\; X(q). 
    \end{eqnarray}
The identity $X(q)=x(q)$
shows that the dynamical susceptibility tends to the 
constrained equilibrium susceptibility at large times. 
The result is simple and it has a clear interpretation. 
It means that asymptotically, during aging, given the configuration 
reached at a time $t_w$ partial equilibrium within the subspace with $C(t,t_w)=q$ is essentially 
achieved before the system can relax to $q-\delta q$.


    
    
    This deep property suggests that the reparametrization invariant part of the dynamics 
    can be described through a fictive quasi-equilibrium process where the physical time 
    is completely elimitated. The system is supposed to be in equilbrium on a correlation 
    scale $q$ and to evolve to a scale $q^-$ choosing configurations according 
    to the Gibbs distribution. 
    Formally this correspond to the Markov chain defined by the following transition matrix
    \begin{eqnarray}
     M_q({\bf S'}|{\bf S}) =\frac{1}{Z_{q^-}[{\bf S}]} \exp(-\beta H[{\bf S'}])\delta(Q({\bf S'},{\bf S})-q^-)
     \label{BPD}
    \end{eqnarray}
   where $Z_{q^-}[{\bf S}]$ is a restricted partition function normalizing the probability.
   The chain, that was called Boltzmann pseudo-dynamics in \cite{franz2013quasi,franz2015quasi} does not correspond to any physical dynamics, 
   it implies macroscopic jumps from one configuration to the next one without solution of continuity.
   In \cite{franz2013quasi,franz2015quasi} however it was suggested that in a system with aging if one chooses $q=q_{EA}$ 
   this fictitious dynamical process provides a coarse-grained reparamtrization invariant effective
   representation of the slow part of the relaxation below $q_{EA}$ of true dynamics: the system fully 
   equilibrates in 'dynamical quasi-states' \cite{franz2000quasi} of amplitude $q_{EA}$ before relaxing below this value; in this relaxation 
   the system explores the configuration space choosing
   new quasi-states at random with Boltzmann probability {\it among the available ones}, i.e. those lying at overlap $q_{EA}$. 
   It is clear from the considerations of the previous section that thanks to Ultrametricity 
   we can go further with the effective description and instead of coarse graining only the short times corresponding to $q_{EA}$, we could choose 
   any arbitrary skeleton value of $q\in Supp(P)$, and the corresponding pseudo-dynamics (\ref{BPD}) would just provide a reparametrization invariant 
   representation of the relaxation below $q$ with the same physical content as the true dynamics. 

\section{Two, perhaps three finite-dimensional puzzles}

\subsection{Is one-step RSB unstable in finite dimensions?}

As we argued above, a system in finite dimensions must be reparametrization-soft and marginally stable, otherwise 
two systems with similar timescales and different effective temperatures would be unable 
to become a two-step RSB when weakly coupled.  Analogously, a 1RSB system coupled with a fullRSB one
must develop a continuum of effective temperatures and hierarchical time scales to become fullRSB itself.

In the mean-field case, however, the equilibrium of a one-step 
systems is not marginal. The situation is analogous to the ferromagnetic case, which is not marginal within 
mean-field (there is an extensive barrier between minima), while it becomes so in finite dimensions, 
being unstable with respect to a small uniform field~\footnote{This however does not give rise to 
divergent susceptibility, as the free-energy has only an essential singularity in $h=0$, and $m(h)=\sign(h)m^+ +h\chi$.} A hypothetical dynamical 1RSB state in finite dimension should be at least 
marginally stable against continuous RSB, a situation that could be called weak fullRSB. Such a situation 
would seem  rather unnatural in a disordered system, by its nature heterogeneous, where the effective  
temperature would have spatial fluctuations.  {It is thus natural to conjecture that close to equilibrium, 
one-step, or more  generally discontinuous solutions are unstable against 
higher RSB, thus leading to a situation where all the values of the correlations except the ones corresponding to the 
trivial equilibrium regime are 'skeleton' values and full 
ultrametricity holds.
}


\subsection{Dynamic timescales in three and four dimensions}

The experimental evidence of spin-glasses seems to point to {\em only two timescales} \cite{berthier2000dynamic,vincent1997slow}, and a non-trivial $P(q)$ \cite{herisson2002fluctuation} with what appears to
be many temperatures -- i.e. the $\chi$ vs $C$ plot appears to have a curved section, apparently violating the Multithermalization. This however should be taken with a grain of salt as it could be preasymptotic effect. Perhaps a clarifying step would be
to see whether the correlation decay in times of hours and days is fitted with the same time-dependence of
the one in microseconds, a comparison that has not, to our knowledge, been done. 
The numerical evidence in three dimensions seems to point to overlap equivalence between site and link overlap \cite{contucci2006overlap}, although we know that this is not possible if there is only one slow timescale and
many temperatures. The probable explanation is that the violation is, however,  too small to be observed numerically.

In four dimensions the situation is much more clear: the little evidence we have \cite{stariolo2001dynamic} seems to point resolutely to dynamic ultrametricity with many scales,  just as in the Sherrington-Kirkpatrick model.

\subsection{Aging strongly far from equilibrium}

Structural glasses as we know them in the laboratory find themselves in extremely long lived metastable states. 
The internal energy density, specific volume etc. persist to off-equilibrium values high above the equilibrium ones for at least geological times. Yet these 
systems slowly age. The response during physical aging in such conditions verifies to an excellent approximation scaling laws with a single effective time scale \cite{struik1977physical}.   Numerical simulations of model glasses
indicate that a description in terms of a single effective temperature, reminiscent of Mean-field 1RSB systems 
is appropriate \cite{parisi1997off,barrat1999fluctuation}. One can ask 

One can conjecture that in such conditions, which are formally very far from the asymptotic situation needed
to have quasi-equilibrium sampling and reparametrization invariance, the dynamics in the aging regime can still be 
approximately described in term of these concepts, as a quasi-equilibrium process where degrees of freedom that
evolve on the same scales are in mutual `multithermalized' equilibrium,
{and that a quasi-equilibrium process as in \cite{franz2013quasi,franz2015quasi} with an properly chosen $q$ would still be an appropriate quasi-reparameterization invariant 
description of the slow dynamics. 
}
\section{Conclusions}

 We have discussed the deep connection between dynamic and equilibrium theoretical constructions as it is realized whenever
during slow dynamics bulk expectation values are close to equilibrium, as it should asymptotically be the case in finite dimensional systems. 
In that case Linear Response Theory allows one to relate dynamical response functions to equilibrium correlations.
 Linear Response  in disordered systems is, however, subtle.
 
The main elements at play are the softness displayed by these systems both in equilibrium and off-equilibrium  with respect to random perturbations, and somewhat paradoxically, the stability of appropriately defined responses and correlations with respect to the same perturbations.

It seems fair to say that the evolution of the subject has been from an almost miraculous ansatz to the gradual understanding of the questions it raises on more robust and method-independent properties.

\vspace{1cm}

   {\bf Acknowledgments} 
    We acknowledge the  support of the Simons Foundation,  Grants No 454941 S. Franz  No 454943 J. Kurchan.



\bibliography{franz_kurchan.bib}

\end{document}